\documentstyle[preprint,aps]{revtex}

\begin{document}
\tighten
\draft

\title{Probing the nature of the seesaw in renormalizable SO(10)}

\author{Borut Bajc$^{(1)}$, Goran Senjanovi\'c$^{(2)}$
and Francesco Vissani$^{(3)}$ }

\address{$^{(1)}${\it J. Stefan Institute, 1001 Ljubljana, Slovenia}}

\address{$^{(2)}${\it International Centre for Theoretical Physics,
34100 Trieste, Italy }}

\address{$^{(3)}${\it INFN, Laboratori Nazionali del Gran Sasso,
Theory Group, Italy}}

\date{\today}
\maketitle

\begin{abstract}
We study the nature of the see-saw mechanism in the context of 
renormalizable SO(10) with 
Higgs fields in the 10-plets and $\overline{126}$-plet 
representations, paying special attention to the 
supersymmetric case. We discuss analytically 
the situation for the second and third generations of fermions 
ignoring any CP violating phase. It is 
shown that $b-\tau$ unification and large atmospheric mixing angle 
strongly disfavor the dominance of the type I see-saw. 
\end{abstract}

\section{Introduction}

We have shown recently \cite{Bajc:2002iw} (see also \cite{Bajc:2001fe}), 
by studying the second and third generations of fermions in the context 
of the minimal renormalizable SO(10) 
theory, that the so-called type II see-saw mechanism naturally 
connects $b-\tau$ unification with the large atmospheric mixing 
angle ($\theta_{atm}$). Subsequent numerical studies for the full 
three generations case \cite{Goh:2003sy,Goh:2003hf}
further enhance the type II case and lead 
to an interesting prediction of a large 1-3 leptonic mixing angle, 
sitting on the experimental limit. 
Similar numerical studies \cite{Matsuda:2001bg,Fukuyama:2002ch} 
also show the same result for the type I see-saw, if one fine-tunes 
the CP phases (see however \cite{Matsuda:2004bq}). In this case, 
though, the connection between $b-\tau$ unification and large 
$\theta_{atm}$ is lost.

At this point one may believe that the issue is closed. However we think 
that analytical results are extremely important since they provide insight 
in this often obscure issue. This was the spirit of our original work 
\cite{Bajc:2002iw} and this is what forced us to work with $2^{nd}$ 
and $3^{rd}$ generations only and to ignore CP violating phases. In 
\cite{Bajc:2002iw} an important question remained unanswered: what 
about type I (canonical) see-saw? We show here how in 
the same context (SO(10) with renormalizable interactions only) 
the experimental facts of approximate $b-\tau$ 
unification and large $\theta_{atm}$ strongly disfavor the 
dominance of type I see-saw in the case of real couplings and 
vacuum expectation values (vevs). 

Before proceeding with our analysis, we briefly review the grand 
unified theory in question and the types of the see-saw mechanisms 
present in minimal left-right symmetric theories (such as SO(10)) in 
general. For some recent reviews and more references see for 
example \cite{Bajc:2003ba,Bajc:2003ht,Mohapatra:2003qw,Raby:2004br}. 

\section{Seesaw in SO(10)}

The idea of SO(10) grand unification is rather old \cite{Fritzsch:nn}, 
and even the more specific and appealing supersymmetric grandunified 
SO(10) theory is with us since long time \cite{Aulakh:1982sw,Clark:ai}. 
By now there are many versions of this theory, and even many ``minimal'' 
versions \cite{Babu:1998wi,Blazek:1999ue,Albright:2000sz}. 
In what follows we stick to the renormalizable version only, 
in order to be able to have a predictive theory. Non-renormalizable 
operators 
depend on what lies beyond SO(10) and thus are not under control. It 
was demonstrated recently \cite{Aulakh:2003kg} 
that the minimal such renormalizable theory 
is based on three generations of matter superfields in the spinorial 
16-dimensional representation and the Higgs superfields in 10, 126, 
$\overline{126}$ and 210 representations. The theory is minimal in the 
sense of simplicity of the Lagrangian (although not of the computations, 
see \cite{Aulakh:2002zr,Fukuyama:2004xs,last}) and having the least 
number of parameters, i.e. predictability.

This theory is especially simple and predictive in the Yukawa sector: 
only two sets of Yukawas with only 15 real components. 
Strictly speaking this is all we need, and 
what we mean by `renormalizable SO(10)' is just a 
theory where the Yukawa are due to interactions with Higgs fields
in the 10-plet and $\overline{126}$-plet representations.
The results we are about to present are valid for 
any theory with the same Yukawa sector, no matter how complicated 
the heavy Higgs sector is. What remains 
undetermined, though, is the nature of the see-saw mechanism 
\cite{Mohapatra:1979ia}. As is well 
known, in any renormalizable left-right symmetric theory, such as for 
example SO(10), there are two different sources of see-saw 
\cite{Lazarides:1980nt,Mohapatra:1980yp}. The first 
is the canonical one, called type I, which takes place through the 
necessary presence of heavy right-handed neutrinos. The right-handed 
neutrinos get their masses through the SU(2)$_R$ triplet $\Delta_R$, 
and L-R symmetry implies the existence of its left-handed counterpart, 
the SU(2)$_L$ triplet $\Delta_L$. In a generic ordinary field theory it 
can be shown \cite{Lazarides:1980nt,Mohapatra:1980yp} 
that $\langle\Delta_L\rangle\ne 0$ necessarily, whereas in 
supersymmetry the situation is more delicate 
\cite{Aulakh:1999cd,Melfo:2003xi}. Let us recall briefly the 
salient features.

The fact that $\langle\Delta_L\rangle\ne 0$ can be seen most eloquently 
by studying the one-loop tadpole for $\Delta_L$ with the type I seesaw 
mechanism: it is clearly divergent. In supersymmetry there is a 
compensating diagram which cancels precisely the divergence. 
In other words, there is no a priori argument against 
$\langle\Delta_L\rangle =0$. The point is simple: 
$\langle\Delta_L\rangle\ne 0$ emerges from a potential term 
(in symbolic notation)

\begin{equation}
V=\Delta_L\Phi^2\Delta_R +...\;,
\label{veff}
\end{equation}

\noindent
where $\Phi$ is the SU(2)$_L\times$SU(2)$_R$ bi-doublet field $\Phi (2,2)$ 
with $B-L=0$. At the cubic level there is no such term, so in supersymmetry 
(susy) without any extra fields the above interaction is absent and 
$\langle\Delta_L\rangle =0$. In other words, in the minimal susy L-R or 
Pati-Salam theory one has the canonical, type I see-saw. In order to achieve 
the $\Delta_L\Phi^2\Delta_R$ interaction, one has two possibilities in 
supersymmetry:

(a) the presence of a SU(2)$_L\times$SU(2)$_R$ field $S(3,3)$, so that 

\begin{equation}
W=\Delta_L\Delta_RS+S\Phi^2+MS^2\;.
\end{equation}

\noindent
After integrating out the (heavy) field $S$, (\ref{veff}) emerges. 

(b) the presence of a $B-L$ carrying bi-doublets $X$ and $\overline{X}$, 
and so

\begin{equation}
W=\Phi\Delta_LX+\Phi\Delta_R\overline{X}+MX\overline{X}\;.
\end{equation}

\noindent
Again, (\ref{veff}) is obtained by integrating out the heavy fields 
$X$ and $\overline{X}$. Equivalently, $X$ and $\overline{X}$ could 
pick up vevs, or, more precisely, the light doublets may 
contain linear combinations of $\Phi$ and $X$ and $\overline{X}$ 
bi-doublets. 

One type of minimal renormalizable SO(10) theory contains 54 and 45 
superfields \cite{Lee:1994je,Aulakh:2000sn}, another one 
a single 210 superfield. In the former 
case, 54 contains $S$, whereas in the latter case 210 contains 
$(2,2,10)$ and $(2,2,\overline{10})$ fields $X$ and $\overline{X}$ 
in the Pati-Salam SU(2)$_L\times$SU(2)$_R\times$SU(4)$_C$ notation. 
Thus, in either case type II see-saw is present together with the 
type I, and the resulting physical consequences become rather hard 
to decipher. Is there a way of disentangling the two sources of seesaw? 
Unfortunately, this question can be answered only within a specific 
model, there is not a general answer. 
We are inclined to favor the case of the 
model with 210, for the following reasons. 
First, this reduces the number of fields (compared with 54+45) 
and thus the number of parameters in the Higgs part of the superpotential. 
Second, and more important, through $10_H210_H126_H$ and 
$10_H210_H\overline{126}_H$ couplings in the superpotential, the $(2,2,15)$ 
fields in $126$ and $\overline{126}$ mix with the $(2,2,1)$ field in $10_H$ 
and thus pick up vevs of order $M_Z$ \cite{Lazarides:1980nt,Babu:1992ia}. 
This in turn implies a possibility 
of correct mass relations for the first and second generations of fermions 
a la Georgi-Jarlskog \cite{Georgi:1979df} 
without any further model building.

It has to be stressed however, that what follows does not depend on 
supersymmetry or on the specific model chosen, but only on the 
coupling of the matter $16_F$  to Higgs $10_H$ and $\overline{126}_H$, 
and on the assumption that the relevant bidoublets get a nonzero vev. 

In short, the Yukawa superpotential is given by (in an obvious notation)

\begin{equation}
W=Y_{10}16_F16_F10_H+Y_{126}16_F16_F\overline{126}_H\;.
\end{equation}

The resulting mass matrices take the form

\begin{eqnarray}
\label{mu}
M_U&=&v_{10}^uY_{10}+v_{126}^uY_{126}\;,\\
\label{md}
M_D&=&v_{10}^dY_{10}+v_{126}^dY_{126}\;,\\
\label{mnd}
M_{\nu_D}&=&v_{10}^uY_{10}-3v_{126}^uY_{126}\;,\\
\label{me}
M_E&=&v_{10}^dY_{10}-3v_{126}^dY_{126}\;,\\
\label{mnr}
M_{\nu_R}&=&\langle\Delta_R\rangle Y_{126}\;,\\
\label{mnl}
M_{\nu_L}&=&\langle\Delta_L\rangle Y_{126}\;,
\end{eqnarray}

\noindent
where $M_{\nu_D}$ stands for the neutrino Dirac Matrix, and $M_{\nu_R}$ 
and $M_{\nu_L}$ for direct Majorana mass matrices for right-handed and 
left-handed neutrinos, respectively. 

From (\ref{mu})-(\ref{mnl}) the light neutrino masses, after integrating 
out the right-handed neutrinos become the mixture of the type I and type II 
see-saw 

\begin{equation}
\label{mn}
M_N=-M_{\nu_D}^TM_{\nu_R}^{-1}M_{\nu_D}+M_{\nu_L}\;.
\end{equation}

A priori, both terms are equally important, and the resulting analysis 
becomes messy. In order to simplify the issue, we have recently discussed 
analytically the case of only two generations, the second and the third 
one, assuming type II seesaw. As mentioned at the outset, the type II 
contribution connects naturally the large atmospheric mixing angle with the 
$b-\tau$ unification. 

In this work we address the issue of the comparison of the two sources 
of seesaw in the same context. 

\subsection{Useful definitions}

The fermion masses depend on the composition of the light higgs 
particles, which we parametrize in the following manner. Let us 
decompose the doublet mass matrix $M_H$ as 

\begin{equation}
\label{uhdh}
U_H^TM_HD_H=M_H^d\;,
\end{equation}

\noindent
where the zero-modes Higgs fields are defined as 

\begin{equation}
H_u=(U_H^\dagger H_{+1})_1\;,\;H_d=(D_H^\dagger H_{-1})_1
\end{equation}

\noindent
and the first two components of $H_Y$ are the weak doublets in 
$10$ and $\overline{126}$ with hypercharge $Y$. 
Thus the expressions of the vacuum expectation values are 

\begin{equation}
v_{10}^X=X^H_{11}v^X\;\;\;,\;\;\; 
v_{126}^X=X^H_{21}v^X\;\;\;,\;\;\;
X=u,d\;\;\;.
\end{equation}

For later use, let us define some derived quantities:

\begin{equation}
\label{xy}
x={U_{21}^HD_{11}^H\over D_{21}^HU_{11}^H}\;\;\;,\;\;\;
y={D_{11}^H\over U_{11}^H}\;\;\;,
\end{equation}

\begin{equation}
\label{alpha}
\alpha=\left(4{U_{11}^HD_{21}^H\over D_{11}^H}\right)^2{v_u^2\over 
\langle\Delta_L\rangle\langle\Delta_R\rangle}\;.
\end{equation}

\section{Can we tell type I from type II?}

Let us begin discussing an important issue: comparing the two forms 
of see-saw is manifestly model dependent, thus one needs a theory 
of fermion masses in order to address it. 
For example, in distinguishing between the 
two types of see-saw one tacitly assumes that they have different 
forms. However, it is possible, at least in principle, that they are 
essentially equivalent. Type II says that $M_N\propto M_{\nu_R}$, 
but if $M_{\nu_D}\propto M_{\nu_R}$, type I would give the same result. 
It appears as too much fine-tuning, but as can be seen from 
(\ref{mu})-(\ref{mnl}) it only requires $v_{10}^u\ll v_{126}^u$. 
If this were to work, it would mean that the whole issue of the 
nature of see-saw is not well defined. However, it can be shown not 
to work in this case. What happens is the following.

Using 

\begin{equation}
(M_U,M_{\nu_D})=v^u(Y_U,Y_{\nu_D})\;\;\;,\;\;\;
(M_D,M_E)=v^d(Y_D,Y_E)\;\;\;,
\end{equation}

\noindent
in (\ref{mu})-(\ref{me}), one obtains a relation among the 
Yukawa couplings of the charged fermions:

\begin{equation}
\label{ylud}
Y_E={1\over 1-x}\left[4yY_U-\left(3+x\right)Y_D\right]\;.
\end{equation}

We are interested in the limiting case $M_{\nu_D}\propto M_{\nu_R}$, 
i.e. $v_{10}^u=0$, or, better $U_{11}^H=0$. From (\ref{xy}) 
this means $x,y=\infty$ with finite $x/y$. (\ref{ylud}) becomes 

\begin{equation}
\label{xyinf}
Y_E=-4\left({y\over x}\right)Y_U+Y_D\;.
\end{equation}

In the 2-generation case this represents 3 equations for only 
two unknowns, $y/x$ and the rotation angle $\theta_D$ in 
$D^TE$, where the unitary matrices $X$ are defined by 
$Y_X=XY_X^dX^T$ (we will assume all model parameters real, 
which gives $X$ orthogonal). We introduce 

\begin{equation}
\epsilon_u={y_c\over y_t}\;\;\;,\;\;\;
\epsilon_d={y_s\over y_b}\;\;\;,\;\;\;
\epsilon_e={y_\mu\over y_\tau}\;\;\;,\;\;\;
\end{equation}

\noindent
and take into account that experimentally 
$\theta_q,\epsilon_i={\cal O}(\delta\approx 10^{-2})$. 
Now we want to see how all this fits into the equation for 
the atmospheric angle. We get it from $M_N\propto Y_d-Y_e$:
first

\begin{equation}
\tan{2\theta_l}={\sin{2\theta_D}\over {y_\tau-y_\mu\over y_b-y_s}
-\cos{2\theta_D}}\;,
\end{equation}

\noindent
and then using (\ref{xyinf}) 

\begin{equation}
\tan{2\theta_l}={\sin{2\theta_q}\over 
{\left(1-\epsilon_u\right)\left[\left(1-\epsilon\right)
\left(1+\epsilon_e\right)-\left(1+\epsilon_d\right)\right]
\over \left(1+\epsilon_u\right)\left(1-\epsilon_d\right)}-
\cos{2\theta_q}}\;.
\end{equation}

The first term in the denominator is ${\cal O}(\delta)$, which 
gives the experimentally unacceptable small atmospheric mixing 
angle $\theta_l=-\theta_q+{\cal O}(\delta^2)$. It means thus, that 
in the minimal theory type I cannot mimic the type II see-saw: 
even if it gives $M_N\propto Y_D-Y_E$, the over-constrained system 
does not allow a large atmospheric solution.

There is an important lesson in this: in this theory 
type I and type II are truly different. However, 
it is clear that the nature of the see-saw mechanism is a meaningful 
question only if one has a theory of fermion masses, i.e. restricted 
Yukawa couplings, otherwise one cannot exclude that
the other version can be made equivalent, meaning that 
we cannot tell type I from type II by neutrino masses alone.  

\section{The 2 generation case (warm-up explorations)}

To illustrate the discussion which follows, 
we start with a simple result for the canonical, type I term. Take 
the 2-3 generation case, and work in the approximation of zero second 
generation masses compared to the third one. It is a straightforward 
exercise to derive the connection between the leptonic (atmospheric) mixing 
angle $\theta_l$ and the quark ($b-c$) mixing angle $\theta_q$, 

\begin{equation}
\label{typeI}
\tan{2\theta_l}={\sin{2\theta_q}\over 2\sin^2{\theta_q}-5/9}\;,
\end{equation}

\noindent
if you assume $b-\tau$ unification (recall that with (\ref{md}) and 
(\ref{me}) it is not automatic). 
Manifestly, a small $V_{cb}$ ($\theta_q\to 0$ limit) 
implies a small $\theta_{atm}$ ($\theta_l\to 0$). Clearly, without a 
possible judicious choice of CP phases 
\cite{Matsuda:2001bg,Fukuyama:2002ch} 
in the 3-generation case (i.e. 
fine-tuning), the type I see-saw is generically strongly disfavored. 

We can do better. Let us study the general case in (\ref{mn}), which 
incorporates both types of see-saw, while still keeping the vanishing 
second generation masses for the sake of illustration. After some 
straightforward computational tedium, one can derive a simple but eloquent 
formula which connects the leptonic and quark mixing angles

\begin{equation}
\label{typeII}
\tan{2\theta_l}={\sin{2\theta_q}\over 2\sin^2{\theta_q}-\Delta}\;,
\end{equation}

\noindent
with

\begin{equation}
\Delta={1\over 1-9\alpha}\left[-5\alpha+\epsilon
\left(1-4\alpha\right)\right]\;,
\end{equation}

\noindent
where 

\begin{equation}
\epsilon={y_b-y_\tau\over y_b}
\end{equation}

\noindent
and $\alpha$ is a relative measure of type I versus type II see-saw, 
defined in terms of the model parameters in (\ref{alpha}). 
In the limit $\alpha\to\infty$ (type I) (\ref{typeI}) is 
reproduced correctly: $\Delta=5/9$ for $y_b=y_\tau$, and so in this 
approximation type I see-saw gives a wrong result of a small $\theta_{atm}$. 
As we mentioned before, it is possible to fine-tune the CP phases in the 
full 3 generations case, but generically speaking there is a problem. On 
the contrary, in the $\alpha=0$ limit (type II), 
$\Delta\approx 1-y_\tau/y_b$ and the large atmospheric mixing is related 
to the $b-\tau$ unification at the GUT scale. 

The type II see-saw emerges in the limit of large 
$\langle\Delta_L\rangle$, i.e. when 
$\langle\Delta_L\rangle\gg v_u^2/\langle\Delta_R\rangle$, 
whereas the type I dominates in the opposite case of small 
$\langle\Delta_L\rangle$. This much is obvious, and formulae 
(\ref{alpha}) and (\ref{uhdh}) give in principle a way of 
quantifying this. Again, in principle, in a complete theory, 
such as the minimal SO(10), it may be possible to determine 
the nature of the see-saw mechanism from the first principles, but in 
practice it is hard. We have unification constraints, and the 
doublet-triplet splitting together with the $d=5$ proton decay limits, 
and this can shed some light in the issue and will be studied in future. 
Here we opt for the bottom-up approach which as we have seen may help 
deciding the nature of the see-saw. 

In this work, we focus on the masses of the second and third 
generations, for which we can show explicit analytical results. 
It is natural to expect that the large atmospheric neutrino mixing 
arises at this level of approximation (while other neutrino 
properties could require to include the first generation masses). 
This is why we will address the question on the nature of the seesaw 
taking advantage  on the observed large atmospheric mixing angle 
$\theta_{atm}$.

\section{The 2 generation case (exact results, no phases)}

At this point one would like to be sure, that the approximation 
with massless second generation quarks and charged leptons is 
at least qualitatively correct. We will now give the general 
formulae and check the conclusions given before. We will see 
that the above approximation is not really needed for the type II 
see-saw to be strongly preferred by data. 

\subsection{General formulae}

Let us start with two matrix equations, which are valid for any 
number of generations as well as for general, complex, parameters.

\begin{eqnarray}
Y_E&=&{1\over 1-x}\left[4yY_U-\left(3+x\right)Y_D\right]\;,\\
Y_N&=&
-\alpha\left[{3\left(1-x\right)Y_D+\left(1+3x\right)Y_E\over 4}\right]
\left(Y_D-Y_E\right)^{-1}
\left[{3\left(1-x\right)Y_D+\left(1+3x\right)Y_E\over 4}\right]\nonumber\\
&+&\left(Y_D-Y_E\right)\;,
\end{eqnarray}

\noindent
where

\begin{equation}
Y_N={4D_{21}^H\over\langle\Delta_L\rangle}M_N
\end{equation}

\noindent
is a dimensionless matrix proportional to the light neutrino matrix 
(\ref{mn}). It is the most important quantity of this paper.

Any symmetric matrix $Y_X$ gets diagonalized by a unitary matrix 
$X$ as ($X=U,D,E,N$)

\begin{equation}
Y_X=XY_X^dX^T\;.
\end{equation}

After defining the unitary matrices

\begin{equation}
V_q=D^\dagger U\;\;\;,\;\;\;
V_l=E^\dagger N\;\;\;,\;\;\;
V_D=D^\dagger E\;\;\;,
\end{equation}

\noindent
we get the equations

\begin{eqnarray}
\label{vdye}
V_DY_E^dV_D^T&=&{1\over 1-x}
\left[4yV_qY_U^dV_q^T-\left(3+x\right)Y_D^d\right]\;,\\
\label{vlyn}
V_lY_N^dV_l^T&=&
-\alpha\left[{3\left(1-x\right)V_D^\dagger Y_D^dV_D^*+
\left(1+3x\right)Y_E^d\over 4}\right]
\left(V_D^\dagger Y_D^dV_D^*-Y_E^d\right)^{-1}\nonumber\\
&&\times\left[{3\left(1-x\right)V_D^\dagger Y_D^dV_D^*+
\left(1+3x\right)Y_E^d\over 4}\right]
+\left(V_D^\dagger Y_D^dV_D^*-Y_E^d\right)\nonumber\\
&=&\left[1-\alpha{9\left(1-x\right)^2\over 16}\right]
\left(V_D^\dagger Y_D^dV_D^*-Y_E^d\right)
-\alpha{3\left(1-x\right)\over 2}Y_E^d\nonumber\\
&&-\alpha Y_E^d\left(V_D^\dagger Y_D^dV_D^*-Y_E^d\right)^{-1}Y_E^d\;.
\end{eqnarray}

It has to be noticed, that in the general complex case $V_q$ is not 
the CKM matrix, since it contains all the phases, i.e. it has as a 
general $N_f\times N_f$ unitary matrix $N_f^2$ real parameters - 
$N_f(N_f-1)/2$ angles and $N_f(N_f+1)/2$ phases. 

We will however consider only the case of real parameters, i.e. no 
CP violating phases. In this case $V_q$ is the truncated CKM matrix
for the second and third generation ($s_q=\sin{\theta_q}$, 
$c_q=\cos{\theta_q}$):

\begin{eqnarray}
V_q=
\pmatrix{
   c_q
&  s_q
\cr   
  -s_q
&  c_q
\cr   }\;,
\end{eqnarray}

\noindent
and similarly ($s_D=\sin{\theta_D}$, 
$c_D=\cos{\theta_D}$):

\begin{eqnarray}
V_D=
\pmatrix{
   c_D
&  s_D
\cr   
  -s_D
&  c_D
\cr   }\;.
\end{eqnarray}

Solving (\ref{vdye}) it is easy to get ($t_q=\tan{\theta_q}$, 
$t_D=\tan{\theta_D}$)

\begin{equation}
\label{x}
{1-x\over 4}=
{\epsilon_d\left(1+\epsilon_ut_q^2\right)
-\left(\epsilon_u+t_q^2\right)\over
\left[\left(1-\epsilon\right)
{1+\epsilon_et_D^2\over 1+t_D^2}-1\right]
\left(\epsilon_u+t_q^2\right)-
\left[\left(1-\epsilon\right)
{\epsilon_e+t_D^2\over 1+t_D^2}-\epsilon_d\right]
\left(1+\epsilon_ut_q^2\right)}\nonumber
\end{equation}

\noindent
and

\begin{eqnarray}
\label{td}
t_D&=&{\left(1-\epsilon_e\right)
\left[\left(\epsilon_u-\epsilon_d\right)+
\left(1-\epsilon_u\epsilon_d\right)t_q^2\right]\over 
2\left(1-\epsilon_u\right)\left(1-\epsilon_e\epsilon_d\right)
t_q}\times\nonumber\\
&\times&\left[1+\sigma\left(1-
{4\left(1-\epsilon_u\right)^2
\left(1-\epsilon_e\epsilon_d\right)\left(\epsilon_e-\epsilon_d\right)
t_q^2\over 
\left(1-\epsilon_e\right)^2
\left[\left(\epsilon_u-\epsilon_d\right)+
\left(1-\epsilon_u\epsilon_d\right)t_q^2\right]^2}\right)^{1/2}\right]
\;\;,\;\;\sigma=\pm 1\;\;.
\end{eqnarray}

Finally, the expression for the leptonic mixing angle is 

\begin{equation}
\label{thetal}
\tan{2\theta_l}={\sin{2\theta_D}
\left(1+
{\alpha D_1\over 1-{9\alpha\over 16}\left(1-x\right)^2}
\right)\over 
{\left(1-\epsilon\right)\left(1-\epsilon_e\right)\over
\left(1-\epsilon_d\right)}-\cos{2\theta_D}+
{\alpha D_2\over 1-{9\alpha\over 16}\left(1-x\right)^2}}\;,
\end{equation}

\noindent
with

\begin{eqnarray}
\label{d1}
D_1&=&{\epsilon_e(1-\epsilon)^2\over
\left[{\epsilon_d+t_D^2\over 1+t_D^2}-\epsilon_e(1-\epsilon)\right]
\left[{1+\epsilon_dt_D^2\over 1+t_D^2}-(1-\epsilon)\right]-
{(1-\epsilon_d)^2t_D^2\over (1+t_D^2)^2}}\;,\\
\label{d2}
D_2&=&3{(1-x)\over 2}
{(1-\epsilon_e)(1-\epsilon)\over (1-\epsilon_d)}\\
&+&{(1-\epsilon)^2\over (1-\epsilon_d)}\times
{\left[{\epsilon_d+t_D^2\over 1+t_D^2}-\epsilon_e(1-\epsilon)\right]
-\epsilon_e^2\left[{1+\epsilon_dt_D^2\over 1+t_D^2}-(1-\epsilon)\right]\over 
\left[{\epsilon_d+t_D^2\over 1+t_D^2}-\epsilon_e(1-\epsilon)\right]
\left[{1+\epsilon_dt_D^2\over 1+t_D^2}-(1-\epsilon)\right]-
{(1-\epsilon_d)^2t_D^2\over (1+t_D^2)^2}}\;.\nonumber
\end{eqnarray}

Equations (\ref{td})-(\ref{d2}) are our main results. In order 
to illustrate their meaning we will solve them in 
leading order of the small parameter 

\begin{equation}
\delta={\cal O}(\epsilon_i,\theta_q)\;.
\end{equation}

The general procedure is the following: choose one of the two solutions 
($\sigma=-1$ or $\sigma=+1$) for $t_D$ in (\ref{td}), then use it in 
(\ref{x}), (\ref{d1}), (\ref{d2}) and finally in (\ref{thetal}).

\subsection{First solution: $\sigma=-1$}

Let's start with the solution with the lower sign:

\begin{eqnarray}
t_D&\approx&{\epsilon_e-\epsilon_d\over 
\epsilon_u-\epsilon_d}t_q={\cal O}(\delta)\;,\\
{1-x\over 4}&\approx&
{(\epsilon_u-\epsilon_d)\over 
(\epsilon_e-\epsilon_d)+\epsilon(\epsilon_u-\epsilon_e)}
={\cal O}(1)\;,\\
D_1&\approx&-{\epsilon_e(1-\epsilon)^2
\over \left[\epsilon_d-\epsilon_e(1-\epsilon)\right]\epsilon}
={\cal O}\left({1\over\epsilon}\right)\;,\\
D_2&\approx&6{(\epsilon_u-\epsilon_d)(1-\epsilon)\over 
(\epsilon_e-\epsilon_d)+\epsilon(\epsilon_u-\epsilon_e)}+
{(1-\epsilon)^2\over \epsilon}={\cal O}\left({1\over\epsilon}\right)\;.
\end{eqnarray}

It is not difficult to show that the only hope for $\theta_l$ 
to be large is that $|\alpha|\le{\cal O}(\delta^2)$ {\it and} 
$|\epsilon|\le{\cal O}(\delta)$. So in this case type I see-saw alone 
seems to be excluded, while $b-\tau$ unification looks also crucial. 
This solution is reminiscent of the original one \cite{Bajc:2002iw} 
with small $\theta_D$ and $b-\tau$ unification needed.

\subsection{Second solution: $\sigma=+1$}

Let us now consider the second solution (\ref{td}):

\begin{eqnarray}
t_D&\approx&{\epsilon_u-\epsilon_d\over t_q}={\cal O}(1)\;,\\
{1-x\over 4}&\approx&
{(\epsilon_u-\epsilon_d)
\over (1-\epsilon)s_D^2}={\cal O}(\delta)\;,\\
D_1&\approx&-{\epsilon_e(1-\epsilon)
\over s_D^2}={\cal O}(\delta)\;,\\
D_2&\approx&-(1-\epsilon)={\cal O}(1)\;.
\end{eqnarray}

Again it is not difficult to show that for $|\alpha|\le{\cal O}(1)$ the 
leptonic mixing angle is ${\cal O}(1)$, and can even be maximal for 
some fine-tuned value of $\alpha$. In general however, pure 
type I see-saw ($\alpha\to\infty$) can be excluded since it gives a 
small $\theta_l={\cal O}(\delta)$. This result is independent on 
the value of $\epsilon$, i.e. on $b-\tau$ unification. 
In the limit $y_{s,\mu}=0$ this solution goes to the $\theta_D=0$ 
solution of \cite{Bajc:2002iw}, which was found out to be unphysical 
in the case of just two bidoublets (but the solution here is physical, 
first because here we have more than two Higgs bidoublets and second 
because the limit $y_{s,\mu}=0$ itself is not physical). 

We see thus that on top of the old solution, we have a new one with 
a bigger parameter space allowed ($|\alpha|\le{\cal O}(1)$) and no need 
for $b-\tau$ unification. 

\subsection{Discussion}

Whether type I or type II dominates can be seen also directly from 
the matrices. The starting point is the right-hand-side of (\ref{vlyn}), 
which looks like 

\begin{equation}
\label{omminus}
-\alpha
\pmatrix{
   \delta
&  \delta/\epsilon
\cr   
   \delta/\epsilon
&  1/\epsilon
\cr   }+
\pmatrix{
   \delta
&  \delta
\cr   
   \delta
&  \epsilon
\cr   }
\end{equation}

\noindent
in the $\sigma=-1$ case, and 

\begin{equation}
\label{omplus}
-\alpha
\pmatrix{
   \delta^2
&  \delta
\cr   
   \delta
&  1
\cr   }+
\pmatrix{
   1
&  1
\cr   
   1
&  1
\cr   }
\end{equation}

\noindent
in the $\sigma=+1$ case. Each element in the matrices gives just 
the order of magnitude and should be thus thought as multiplied 
with a coefficient of order one.

It is now immediately clear that in the $\sigma=-1$ case type 
II see-saw dominates (all the single matrix elements) if 
$|\alpha|< {\cal O}(\epsilon^2)$ and $|\epsilon|\le{\cal O}(\delta)$, 
because then each element of the first (type I) matrix is smaller 
than the corresponding element of the second (type II) matrix.  
In the $\sigma=+1$ case, type II see-saw dominates 
as soon as $|\alpha|\le {\cal O}(\delta)$. But even in the partially 
mixed case, i.e. when ${\cal O}(\delta)<|\alpha|\le{\cal O}(1)$ the sum in 
(\ref{omplus}) equals just the second term and gives a large 
(${\cal O}(1)$) atmospheric mixing angle. 
And this is again the case with a large mixing angle, but 
also in the mixed type I plus type II case with 
$\alpha={\cal O}(\delta^2)$ the angle is already large. 

So to summarize, if type II see-saw dominates, then the atmospheric 
mixing angle is large in the $\sigma=+1$ case, but in the $\sigma=-1$ 
case this happens only if $b-\tau$ unification is realized. 
The opposite is not always 
true however: if the angle is large, type II does not need to completely 
dominate. For a range of $\alpha$ the $33$ element of type I can be 
of the same order as the $33$ element of type II and still give a 
large angle (again, $b-\tau$ unification has to hold if $\sigma=-1$).

\section{Summary and outlook}

The see-saw mechanism, it is often said, provides a simple and natural 
rationale for the smallness of neutrino mass. The real issue, though, 
is how to test it experimentally. In principle 
the answer is that simple: find the 
right-handed neutrino or a heavy SU(2)$_L$ triplet, whatever be 
its source, and find the corresponding couplings with light neutrinos. 
The trouble, as we all know, is that we expect the see-saw scale to 
be large, above $10^{10}$ GeV or so, and thus we are left with 
indirect consequences only. This is reminiscent of any high 
energy physics, such as proton decay in GUTs.

In this paper we addressed the issue of the nature of the see-saw 
mechanism in a context of a well defined theory, with a predictive 
low energy effective theory: 
a renormalizable SO(10) theory, with higgs fields in 
the 10-plet and $\overline{126}$-plet representations. In view of 
theoretical motivations, we discussed in more detail supersymmetric 
models but the results are of more general validity. 
We studied analytically the issue of the second and third 
generations only. We had already noticed that the type II see-saw, 
based on the heavy SU(2)$_L$ triplet fits nicely: $b-\tau$ 
unification automatically implies a large atmospheric mixing angle. 

Here we completed the program of investigation: 
we find that in no case type I can 
be a dominant source of neutrino masses, unless one fine-tunes the 
CP violating phases \cite{Matsuda:2001bg,Fukuyama:2002ch} 
(but see \cite{Fukuyama:2002ch}). 
However, type I can be present and even compete with type II. 
More work will be needed before one can disentangle, if ever, 
the two sources of the see-saw mechanism.

What about including the CP phases in this analytic work? We 
have made an attempt to do that, but the amount of computational 
tedium wipes off any transparency. The point is a number of 
unmeasured phases beyond $\delta_{CKM}$, which cannot be 
rotated away at the large scale where one is utilizing the 
SO(10) structure. Some of these phases could be 
measured one day (?) in proton decay factories, but, today, 
this is simply science fiction.

\acknowledgments

We are grateful to Pavel Fileviez Perez and Alejandra Melfo for 
useful comments and a careful reading of the manuscript.
The work of B.B. is supported by the Ministry of Education, Science
and Sport of the Republic of Slovenia.
The work of G.S. is partially supported by EEC, under
the TMR contracts ERBFMRX-CT960090 and HPRN-CT-2000-00152.
We express our gratitude to INFN, which
permitted the development of the present study
by supporting an exchange program with the
International Centre for Theoretical Physics.

{\bf Note Added:}

Yesterday, a new paper that considers related issue appeared 
\cite{mohapatra}. Where it is possible to compare, 
their results agree with ours. 
They investigate numerically the $3\times 3$ 
complex case and conclude that a renormalizable model cannot account 
for a positive $\rho$ in $V_{CKM}$. 
We would like to point out that they
assume (their eq. (5) and (6)) that the light Higgs doublets 
live only in $10$ and $\overline{126}$, while in general this is 
not necessarily the case. It would be interesting also to see, 
whether the different possibilities for $\theta_D$ (our cases 
$\sigma=+1$ and $\sigma=-1$) have been taken into account and analyzed.

\end{document}